\documentclass[twocolumn,floatfix,eqsecnum,prb,superscriptaddress]{revtex4}

\usepackage{graphicx}
\usepackage{dcolumn}
\usepackage{bm}
\usepackage{lipsum}
\usepackage{mathtools}
\usepackage{braket}
\usepackage[caption=false]{subfig}
\usepackage{xcolor}
\usepackage{amssymb}

\begin{document}


\title{Magnetization Dynamics and Peierls Instability in Topological Josephson Structures}

\author{Adrian Reich}  
\affiliation{Institute for Theoretical Condensed Matter Physics, Karlsruhe Institute of Technology, 76131 Karlsruhe, Germany}

\author{Erez Berg}
\affiliation{Department of Condensed Matter Physics, Weizmann Insitute of Science, Rehovot 76100, Israel}

\author{Jörg Schmalian}
\author{Alexander Shnirman}
\affiliation{Institute for Theoretical Condensed Matter Physics, Karlsruhe Institute of Technology, 76131 Karlsruhe, Germany}
\affiliation{Institute for Quantum Materials and Technologies, Karlsruhe Institute of Technology, 76344 Eggenstein-Leopoldshafen, Germany}

\date{\today}

\begin{abstract}
We study a long topological Josephson junction with a ferromagnetic strip between two superconductors. The low-energy theory exhibits a non-local in time and space interaction between chiral Majorana fermions, mediated by the magnonic excitations in the ferromagnet. While short ranged interactions turn out to be irrelevant by power counting, we show that sufficiently strong and  long-ranged interactions may induce a $\mathbb{Z}_2$-symmetry breaking. This spontaneous breaking leads to a tilting of the magnetization perpendicular to the Majorana propagation direction and the opening of a fermionic gap (Majorana mass). It is analogous to the Peierls instability in the commensurate Fr\"ohlich model and reflects the nontrivial interplay between Majorana modes and magnetization dynamics. Within a Gaussian fluctuation analysis, we estimate critical values for the temporal and spatial non-locality of the interaction, beyond which the symmetry breaking is stable at zero temperature -- despite the effective one-dimensionality of the model. We conclude that non-locality, i.e., the stiffness of the magnetization in space and time, stabilizes the symmetry breaking. In the stabilized regime, we expect the current-phase relation to exhibit an experimentally accessible discontinuous jump. At nonzero temperatures, as usual in the 1D Ising model, the long-range order is destroyed by solitonic excitations, which in our case carry each a Majorana zero mode. In order to estimate the correlation length, we investigate the solitons within a self-consistent mean-field approach.
\end{abstract}

\maketitle


\section{Introduction}

The surface of a three-dimensional strong topological insulator (TI) has been predicted to host zero- and one-dimensional Majorana modes when gapped by a superconducting or magnetic covering \cite{fu_superconducting_2008,fu_probing_2009}. A geometry of particular significance in this context is the topological Josephson junction, comprising two superconductor-covered areas of the TI surface, separated by an uncovered or ferromagnetic strip.

There exists a variety of works revolving around such systems, examining for example the role of the Majorana and other bound states concerning the current-phase relation of the Josephson current \cite{tanaka_manipulation_2009,potter_anomalous_2013,hegde_topological_2020,backens_current--phase_2021}. In the case of a ferromagnet being deposited in the junction, possibilities to manipulate the properties of the supercurrent and quasiparticle states by means of the magnetization are of interest as well \cite{tanaka_manipulation_2009,zyuzin_josephson_2016,bobkova_magnetoelectrics_2016,amundsen_vortex_2018}.

When studying the magnetization’s effect on these electronic properties, one should keep in mind, however, that the magnetization within a ferromagnet is not perfectly rigid but exhibits space and time dynamics, which can be described by the Landau-Lifshitz-Gilbert (LLG) equation \cite{landau_theory_1935,gilbert_formulations_1956,gilbert_phenomenological_2004}. In this context it was predicted that due to the spin-orbit coupling a significant torque can be exerted on the magnetization in both ordinary \cite{konschelle_magnetic_2009} and topological Josephson junctions \cite{nashaat_electrical_2019}. This can lead to precession and reorientation of the magnetization direction.

In particular, in Ref.\ \onlinecite{nashaat_electrical_2019} 
a strong dependence of the Josephson coupling on the 
magnetization was obtained, which leads to an additional effective field in the LLG equations. This, in turn, allowed controlling the magnetization by the Josephson current in the 
voltage driven regime. 

\begin{figure}
    \centering
    \includegraphics[width=0.45\textwidth]{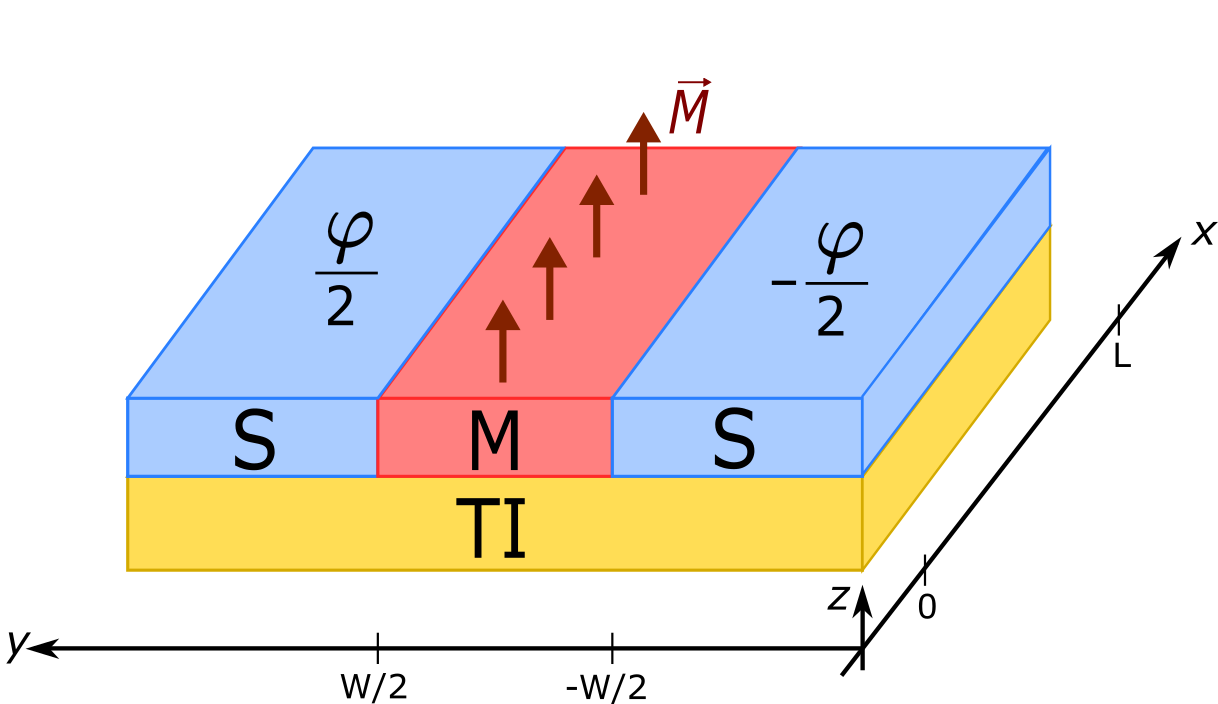}
    \caption{Considered geometry of a superconductor-ferromagnet-superconductor (SMS) junction of width $W$ and length $L$ on the surface of a 3D strong topological insulator (TI) with a phase difference $\varphi$ between the superconductors.}
    \label{fig:model}
\end{figure}

Here, in contrast, we consider the zero-bias case and focus on the coupling between the magnons and the one-dimensional Majorana modes counter-propagating along the junction. We consider both the fermionic picture, in which the magnons are integrated out, and the complementary bosonic picture, in which the fermions are integrated out. 

In particular, we obtain a low-energy effective action describing the coupling between the Majorana fermionic modes and the bosonic magnons, resembling a `Majorana variant' of the Fr\"ohlich model and exhibiting the characteristic Peierls instability in the mean-field approximation\cite{frohlich_theory_1954}. In the present case, this corresponds to a tilting of the magnetization  and the opening of a Majorana mass gap. An analogous broken symmetry has recently been predicted for the relative phase in a Josepshon junction comprised of 2D time-reversal invariant superconductors\cite{ruiz_josephson_2022}. 

Following the fluctuation analysis of the BCS superconductivity \cite{kos_gaussian_2004,fischer_short-distance_2018}, we employ a similar technique and determine a parameter regime in which, within a Gaussian approximation for the fluctuations, the mean-field result is stable at zero temperature. The relevant parameter stabilizing the broken symmetry can be identified as the non-locality of the effective four-Majorana interaction, or, equivalently, the time- or/and space rigidity of the magnons. The long-range order in the system at $T>0$ can then be expected to be destroyed by solitonic excitations, for which the self-consistency problem is examined.

\section{The Model}

Consider a topological superconductor-ferromagnet-superconductor (SMS) junction as depicted in Fig.\ \ref{fig:model}. The $s$-wave superconductors introduce superconducting gaps in the TI surface in the regions they cover due to the proximity effect\cite{alicea_new_2012} and we suppose that the gaps are equal in magnitude $\Delta_0$ on both sides of the junction but differ by a relative phase $\varphi$. The ferromagnetic insulator causes an effective exchange field $\vec{h}_\text{eff}$ in the underlying surface of the TI which couples to the electrons' spin and is proportional to the ferromagnet's magnetization $\vec{h}_{\text{eff}}(\bm{r}) = \alpha\vec{M}(\bm{r})$ with a proportionality constant $\alpha$. Here, 3D vectors are denoted by an arrow above the symbol, 2D vectors are written in bold. 

The Hamiltonian describing this setup is given by $H=\frac{1}{2}\int\text{d}\bm{r}\,\Psi^\dagger h\Psi$ where $\Psi = (\psi_\uparrow,\psi_\downarrow,\psi^\dagger_\downarrow,-\psi^\dagger_\uparrow)^T$ and the BdG Hamiltonian reads
\begin{equation}
\begin{split}
h = -iv_{{\rm F}}&\tau_z\bm{\sigma}\cdot\bm{\nabla} - \mu\tau_z + \alpha \vec{M}(\bm{r})\cdot\vec{\sigma}\\&+ \Delta_0(y)(\cos\varphi(\bm{r})\tau_x+\sin\varphi(\bm{r})\tau_y)\ . \label{eq_hSMS}
\end{split}
\end{equation}
As described above $\Delta_0(y) = \Delta_0\,\Theta(|y|-W/2)$, $\varphi(\bm{r}) = \varphi(x)/2\,[\Theta(y-W/2)-\Theta(-(y+W/2))]$, 
$\vec{M}(\bm{r}) = M_{\rm S}\,\vec{m}(\bm{r})\Theta(W/2-|y|)$
with $v_{\rm F}$ being the Fermi velocity and $\mu$ the chemical potential. The saturation magnetization is denoted by $M_{\rm S}$ and $|\vec{m}|=1$.

The quasiparticle dispersion for the case of a ferromagnet with spatially homogeneous magnetization in $z$-direction $\vec{m}(\bm{r}) = \vec{e}_z$ deposited on a 3D TI reads $
\varepsilon_{\text{TI-M}}(\bm{k}) = \sqrt{v_{\rm F}^2\bm{k}^2+\alpha^2M_{\rm S}^2}\pm\mu$,
which is gapped when $\alpha M_{\rm S}\equiv M>\mu$, i.e. when the Fermi level lies within the mass gap of the Dirac spectrum induced by the exchange field. As shown in 
Ref.~\onlinecite{fu_probing_2009}, a
chiral 1D Majorana mode emerges at the interface between the regions with superconducting and magnetic gaps. In the geometry of Fig.\ \ref{fig:model} two counterpropagating chiral Majorana modes would emerge near each interface. The latter hybridize with an amplitude $\propto \cos(\varphi/2)$ and, thus, decouple at $\varphi = \pi$.

Here we mostly focus on the opposite regime of $M<\mu$. In this case no gap is induced for $|y|<W/2$. For a narrow junction, $W\ll v_{\rm F}/\Delta_0$, and with the phase difference additionally fixed to be $\varphi=\pi$, the situation considered by Fu and Kane in their seminal paper \cite{fu_superconducting_2008} is obtained. The junction becomes a non-chiral Majorana wire with two counter-propagating Majorana modes spread across its whole width. A deviation of the phase difference from a value of $\pi$, $\epsilon=\pi-\varphi$ hybridizes the two Majorana modes, opening up a Majorana mass gap.

Allowing the magnetization direction to slightly deviate from the $z$-axis, we find that the $m_y$-component plays the same role as $\epsilon$ in hybridizing the two Majorana modes with some coupling constant $\lambda$. In Appendix \ref{app_deriving-effham}, the corresponding low-energy effective Hamiltonian is derived. Since we are interested in the interplay between Majorana modes and the magnetization, we fix from now on $\epsilon=0$. Below we also consider what happens when we allow for small deviations from $\varphi=\pi$.

We note that for $\vec{m}=\vec{e}_z$, there is a symmetry present described by the symmetry operator $F = \sigma_z I_y \mathcal{K}$, where $I_y$ denotes $y$-inversion and $\mathcal{K}$ complex conjugation. This symmetry is broken as soon as $\vec{m}$ deviates from the $z$-axis.

For the micromagnetic description of the magnetization dynamics, we introduce a large easy axis anisotropy $B$ in $z$-direction, such that the $x$- and $y$-components of the magnetization can reasonably assumed to be small $\vec{m}=\left(m_x,m_y,1-(m_x^2+m_y^2)/2\right)^T$. Furthermore, we take the junction's width to be small compared to the magnetic coherence length in order for the magnetization direction to only depend on the $x$-coordinate. The magnetic energy shall also include an exchange coupling $A$, such that the corresponding real-time Lagrangian reads
\begin{align}
     \mathcal{L} = &\frac{M_{\rm S}}{2\gamma}(\dot{m}_xm_y-m_x\dot{m}_y)\\&-A\left((\partial_xm_x)^2+(\partial_xm_y)^2\right)-B\left(m_x^2+m_y^2\right),\notag
\end{align}
where the first term corresponds to a Berry phase, ensuring $|\vec{m}|=1$ at all times. The associated equation of motion is the dissipationless LLG equation with gyromagnetic ratio $\gamma$. Note that, due to the restriction of $\vec{m}$ to the unit sphere, there is only one independent degree of freedom in the dynamics of the magnetization, leading to the equal-time commutator of the free quantized bosonic fields $m_x$ and $m_y$ being non-zero $[m_\alpha(x),m_\beta(x^\prime)] = i\frac{\gamma}{M_{\rm S}}\varepsilon_{\alpha\beta}\delta(x-x^\prime)$.\cite{gongyo_effective_2016} In the absence of coupling to the fermions, the dispersion of the magnonic excitations is given by $\omega_q = \frac{2\gamma}{M_{\rm S}}(Aq^2+B)$.

Altogether, the effective Euclidean action reads
\begin{widetext}
\begin{equation}
    S = \int\text{d}\tau\,\text{d}x\,\left[\frac{1}{2}\begin{pmatrix} \gamma_{\rm R} \\ \gamma_{\rm L} \end{pmatrix}^T\begin{pmatrix} \partial_\tau - iv\partial_x & i\lambda m_y \\ -i\lambda m_y & \partial_\tau + iv\partial_x \end{pmatrix}\begin{pmatrix} \gamma_{\rm R} \\ \gamma_{\rm L} \end{pmatrix} + \begin{pmatrix}m_x \\ m_y \end{pmatrix}^T \begin{pmatrix} -A\partial_x^2 + B & i\frac{M_{\rm S}}{2\gamma}\partial_\tau \\ -i\frac{M_{\rm S}}{2\gamma}\partial_\tau & -A\partial_x^2+B  \end{pmatrix}\begin{pmatrix}m_x \\ m_y \end{pmatrix} \right]. \label{eq_ACTION}
\end{equation}
\end{widetext}
At $\lambda=0$ the action of the magnetization dynamics corresponds to a charged scalar with $U(1)$ symmetry. This $U(1)$ invariance reflects the spin rotation invariance in the $x-y$-plane in spin space and prevents spontaneous symmetry breaking. However, the coupling to the fermions, which are governed by spin-orbit interaction, breaks this symmetry down to $\mathbb{Z}_2$, where $\pm m_y$ describe degenerate configurations. Hence, the coupling to fermions may give rise to spontaneous symmetry breaking in the ground state.

Note that although this action has been motivated by a specific system, our considerations below are generalizable to other instances of one-dimensional fermionic modes coupled to a bosonic field, such as the behavior discussed in Ref.~\onlinecite{ruiz_josephson_2022}.

\section{Analysis of the Majorana-magnon interaction}
\subsection{Fermionic picture}
Since, in contrast to the free Majoranas, the spectrum of the magnetic degrees of freedom is gapped, it seems most natural to integrate out the bosons. The magnetic part of the action is diagonalized by introducing the complex scalar field $\phi$ with $m_x=\sqrt{\frac{\gamma}{2M_{\rm S}}}(\phi+\phi^\ast)$ and $m_y=-i\sqrt{\frac{\gamma}{2M_{\rm S}}}(\phi-\phi^\ast)$. One then straightforwardly obtains the effective four-fermion interaction
\begin{eqnarray}
   &&S_{\rm eff}^{\rm int} = -\frac{\lambda^2}{4}\nonumber\\
   &&\times\int\text{d}X\text{d}X^\prime \,\gamma_{\rm L}(X)\gamma_{\rm R}(X)\mathcal{G}_{\rm m}(X,X^\prime)\gamma_{\rm R}(X^\prime)\gamma_{\rm L}(X^\prime)\ ,
   \nonumber\\ \label{eq_4fermion}
\end{eqnarray}
where $X=(x,\tau)$ and $\mathcal{G}_{\rm m}^{-1} = \frac{M_{\rm S}}{2\gamma}\partial_\tau-A\partial_x^2+B$. This interaction is non-local in both space and time with correlation lengths $\xi_{x,\rm{f}}\sim\sqrt{A/B}$ and $\xi_{\tau,\rm{f}}\sim M_{\rm S}/\gamma B$ respectively. For fixed $B$, the correlation length in space is thus governed by the exchange coupling or `stiffness' of the magnetization, while the correlation length in time is proportional to the inverse of the magnonic excitation gap $1/\omega_{q=0}$. If the corresponding correlation lengths are sufficiently small, a gradient expansion of $\mathcal{G}_{\rm m}$ in (\ref{eq_4fermion}) is justified. There, it has to be noted that the zeroth order term, involving no derivatives of the four Majorana fields taken at the same point in space and time, vanishes due to Fermi statistics. The lowest order non-vanishing contributions to the interaction are therefore of the form $\gamma_{\rm R}(\partial_{X_i}\gamma_{\rm R})\gamma_{\rm L}(\partial_{X_j}\gamma_{\rm L})\ (i,j=1,2)$ and can be shown to be highly irrelevant in the RG sense \cite{rahmani_phase_2015}, leaving the Majorana modes gapless and the magnetic degrees of freedom in a disordered phase with vanishingly small correlation length. It is thus clear that in a parameter regime, where $A$ and $M_{\rm S}/\gamma$ are small enough for a gradient expansion to be applicable, the interaction has no qualitative effects on the system. Symmetry breaking is a strong-coupling phenomenon, at least for any finite range interaction. In order to see whether increasing the values of $A$ and/or $M_{\rm S}/\gamma$ leads to a cross-over to a non-trivial phase, it turns out to be advantageous to work in the bosonic picture instead.

\subsection{Peierls instability of the effective bosonic action}

\begin{figure}
    \centering
    \includegraphics[width=0.48\textwidth]{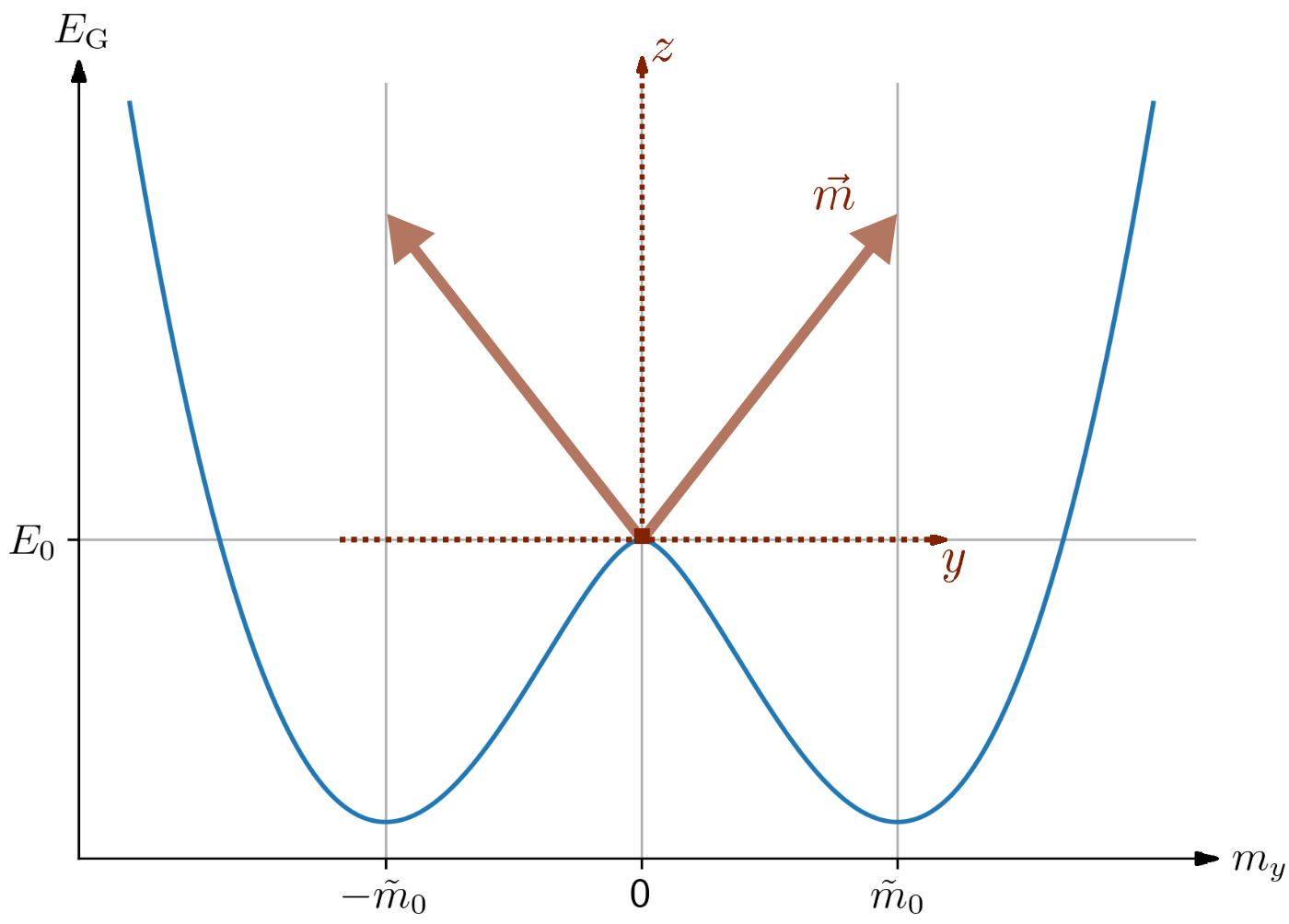}
    \caption{The double well potential arising for $m_y$ in mean-field theory is shown in blue. The red arrows represent the two corresponding symmetry-broken ground state configurations of the magnetic field, tilted away from the $z$-axis with $m_y$ acquiring a finite vacuum expectation value $\braket{m_y}=\pm\tilde{m}_0$.}
    \label{fig:doublewell}
\end{figure}
Let us assume for the moment that the systems breaks the $\mathbb{Z}_2$-symmetry. Then fermions are gapped and can be integrated out.
Integrating out the fermionic degrees of freedom in (\ref{eq_ACTION}) leaves us with the effective bosonic action
\begin{equation}
    S_{\rm eff} = S_{\rm m} - \frac{1}{2}\text{tr}\log\mathcal{G}^{-1}, \label{eq_trlogeffaction}
\end{equation}
where 
\begin{equation}
     \mathcal{G}^{-1} = \begin{pmatrix} \partial_\tau-iv\partial_x & i\lambda m_y \\ -i\lambda m_y & \partial_\tau+iv\partial_x \end{pmatrix}
\end{equation}
and $S_{\rm m}$ is the purely magnetic part of (\ref{eq_ACTION}). 
We assume for the moment that it is permissible to perform this mean field analysis. Below we will analyze fluctuations that go beyond the mean-field approach and discuss the stability of the assumption of long-range order.
For constant $m_y(x,\tau) = \tilde{m}_0$ and within mean field theory,
this action has a saddle point $\delta S/\delta m_y =0$ at $\pm \tilde{m}_0\neq 0$ satisfying the BCS-like gap equation
\begin{equation}
    \frac{2B}{\lambda^2} = \int_0^\Lambda\frac{\text{d}k}{2\pi}\,\frac{\tanh(\sqrt{v^2k^2+\lambda^2\tilde{m}_0^2}/2T)}{\sqrt{v^2k^2+\lambda^2\tilde{m}_0^2}},
\end{equation}
where we took the limit $L\rightarrow\infty$ and introduced a UV momentum cut-off $\Lambda$. 

Thus, at low temperatures, 
\begin{equation}
\lambda \tilde{m}_0\equiv\tilde{\Delta}= 2v\Lambda e^{-4\pi vB/\lambda^2}
\end{equation}
and the corresponding ground state energy in terms of $m_y$ exhibits the characteristic double-well shape $E_{\rm G}/L = [B+\frac{\lambda^2}{4\pi v}(\log|\frac{\lambda m_y}{2v\Lambda}|-\frac{1}{2})]m_y^2+E_0/L$ (see Fig.\ \ref{fig:doublewell}), where $E_0$ is the contribution from $m_y=0$. This spontaneous symmetry breaking at the mean-field level, signifying an instability of the easy axis with $m_y$ acquiring a finite ground state expectation value, can be understood as follows: a positive energy cost, $\propto B m_y^2$, of magnetization deviating from the easy axis direction is balanced by the energy gain, $\propto \frac{\lambda^2}{4\pi v}(\log|\frac{\lambda m_y}{2v\Lambda}|-\frac{1}{2})\,m_y^2$, emerging due to the opening of the fermionic gap.
 It is the Majorana fermion analog to the Peierls instability of the one-dimensional Fröhlich model \cite{frohlich_theory_1954,mckenzie_microscopic_1995}
in the commensurate regime, where the order parameter is real. A real order parameter furthermore means that the broken symmetry is discrete, such that the Mermin-Wagner theorem does not apply. Nevertheless, even if the mean-field solution turns out to be stable, with the system being one-dimensional any emerging long range order has to be expected to be prohibited by the formation of domain walls at $T>0$ in the thermodynamic limit of large $L$. This is due to the fact that the energetic cost of creating domain walls, with $m_y$ switching sign along the junction and interpolating between the two minima, is in 1D always outweighed by the ensuing gain in entropy, as is well-known from the Ising model and Peierls' argument \cite{peierls_isings_1936,griffiths_peierls_1964}. However, drawing the analogy between the Ising model and our system further, at any finite size $L$ the coherence length $\xi$, given by the average distance between two domain walls, can be expected to be exponentially large at low temperatures $\xi \sim b\,e^{E_{\rm DW}/T}$, where $b$ is the characteristic width of a domain wall and $E_{\rm DW}$ its energy, suggesting significantly large stretches of an ordered magnetic phase with hybridization between $\gamma_{\rm R}$ and $\gamma_{\rm L}$ to be realized. This is in contrast to the results we obtained in the gradient expansion of the fermionic picture. 

From this, one can suspect the existence of a cross-over or a phase transition from a phase with small $A$ and $M_{\rm S}/\gamma$, where the gradient expansion is valid and any mean-field considerations in the bosonic picture are rendered unusable due to large fluctuations, to a phase with large $A$ or large $M_{\rm S}/\gamma$, where the fluctuations are suppressed and the mean-field solution is stabilized such that domain walls are a meaningful concept with the interaction in the fermionic picture being very long-ranged. 

Qualitative arguments for analogous mechanisms have already been given in earlier publications. In Refs.\ \onlinecite{lee_conductivity_1974} and \onlinecite{brazovskii_dynamics_1976} devoted to the Peierls instability, the validity of mean-field theory is assumed based on slow response times of the phononic modes. In Ref.\ \onlinecite{gogolin_bosonization_2004}, the authors expect a large temporal stiffness of the bosons mediating the interaction in a Tomonaga-Luttinger liquid to stabilize the symmetry-broken phase (they call this regime ''adiabatic limit''). In the following, we aim to verify and quantify these qualitative considerations on the level of a Gaussian approximation for the fluctuations around the mean-field solution.

\begin{figure*}[ht]
    \centering
    \subfloat[\label{fig:numsolution_gapeq}]{
    \includegraphics[width=0.45\textwidth]{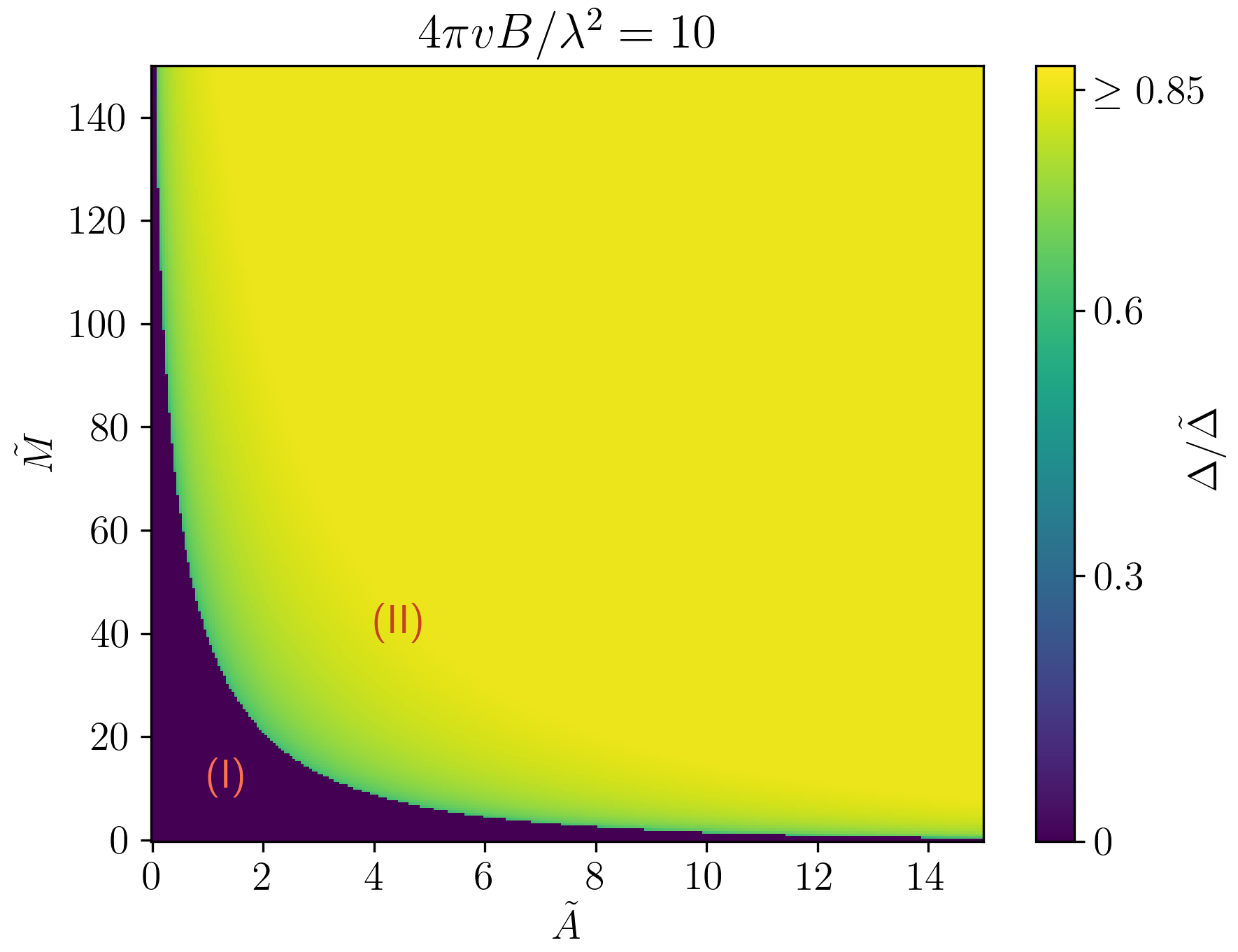}}
    \hspace{0.5cm}
    \subfloat[\label{fig:higgsgap}]{
    \includegraphics[width=0.45\textwidth]{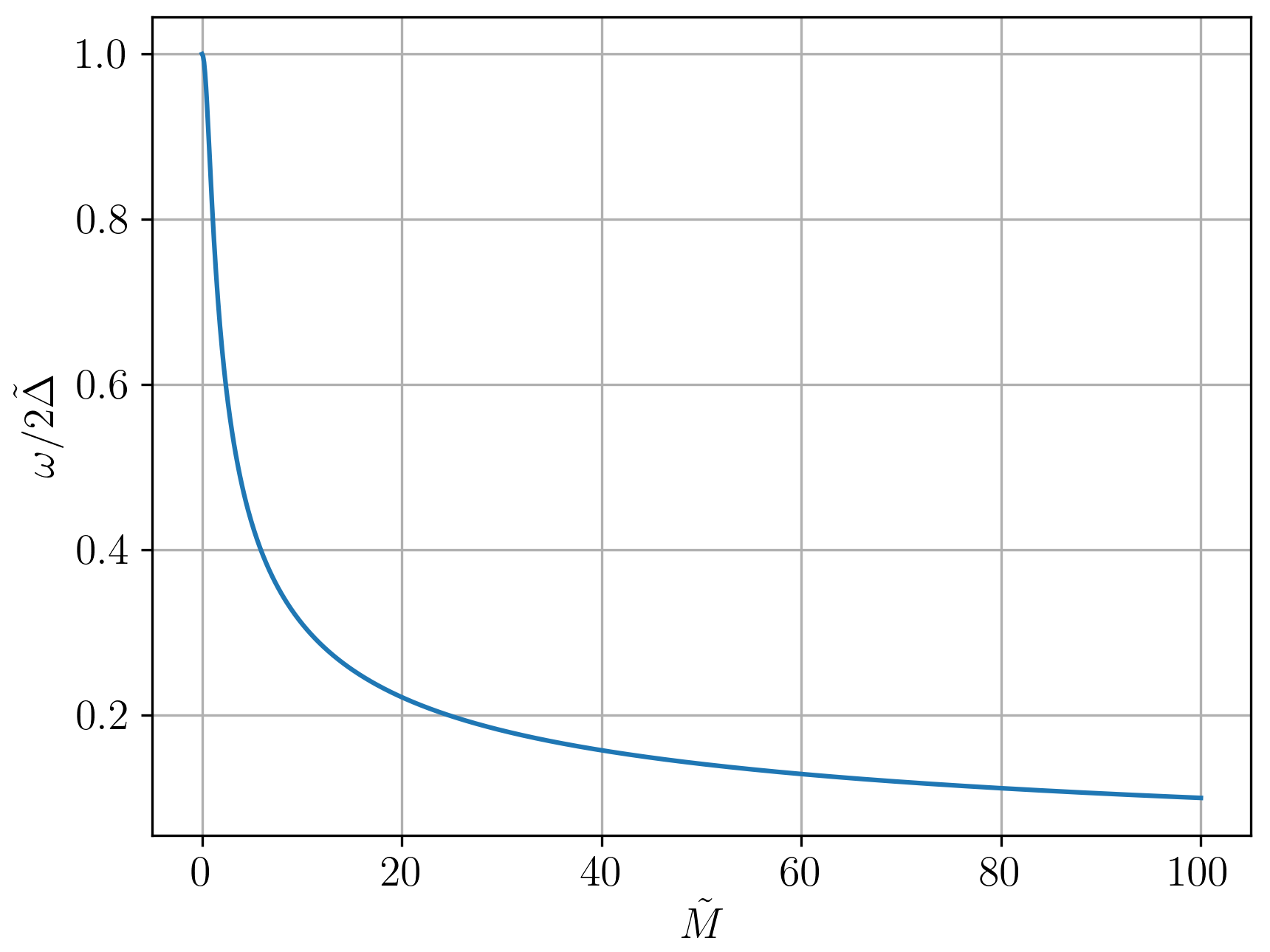}}
    \caption{a) Regions of order (II) with broken Ising symmetry and magnetization canting, perpendicular to the propagation direction, and disorder (I), where fluctuations destroy the ordered state. Shown by color are the solutions $\Delta/\tilde{\Delta}$ to the gap equation (\ref{eq_gapeq}) for $4\pi vB/\lambda^2=10$. $\Delta=0$ corresponds to no solution existing. If $\tilde{A}=0$, a solution $\Delta/\tilde{\Delta}\neq 0$ is obtained for $\tilde{M}\gtrsim 222$ and conversely a value of $\tilde{A}\gtrsim 22$ is needed for a nonzero solution if $\tilde{M}=0$. b) The excitation gap for fluctuations, i.e. the `Higgs gap', in units of $2\tilde{\Delta}$ depending on $\tilde{M}$, where for illustrative purposes we set $\Delta = \tilde{\Delta}$ and thus suppose we are deep in the ordered phase.}
    \label{fig:numsolutions}
\end{figure*}

\subsection{Fluctuation analysis}

Next, we examine the conjectured transition when varying $A$ and $M_{\rm S}/\gamma$ by analyzing the way in which Gaussian fluctuations around a mean-field solution at zero temperature affect the gap equation, as was done for the BCS theory in Refs.~\onlinecite{kos_gaussian_2004} and \onlinecite{fischer_short-distance_2018}. 
Our reasoning is motivated by the following logic: Since symmetry breaking in the Ising universality class is allowed in principle at $T=0$, we assume a finite value of the order parameter. Then, we study Gaussian fluctuations. If they are small, the assumption of order is justified and consistent. If fluctuations are large, they likely destroy long-range order. Then, the above mean-field approach is not justified. While formally this is an uncontrolled approximation to a strong-coupling problem, it gives insights into the parameter regime where canted Ising order is allowed. It gives the order of magnitude of the more microscopic parameters where this order prevails. We should add that performing a numerical density matrix renormalization group calculation for a model in the same universality class does indeed yield the discussed symmetry breaking.\cite{reich_unpublished_2023}



To this end, we expand the action (\ref{eq_trlogeffaction}) around some assumed minimum $m_y=m_0$, which is not necessarily the mean-field minimum $\tilde{m}_0$ from above, up to second order in the fluctuations $(\delta m_x,\delta m_y)$ around it, $m_y(x,\tau)=m_0+\delta m_y(x,\tau)$ and $m_x(x,\tau)=\delta m_x(x,\tau)$, to find
\begin{equation}
\begin{split}
    S_{\rm eff} = \frac{T}{L}\sum_{q,\omega_m}\begin{pmatrix}\delta m_x \\ \delta m_y\end{pmatrix}^T  &\mathcal{D}_{q,\omega_m}^{-1}\begin{pmatrix}\delta m_x \\ \delta m_y\end{pmatrix} \\ &+ \frac{L}{T}Bm_0^2-\frac{1}{2}\text{tr}\log \mathcal{G}_0^{-1},
\end{split}
\end{equation}
where
\begin{equation}
    \mathcal{D}_{q,\omega_m}^{-1}=\begin{pmatrix} Aq^2+B & \frac{M_{\rm S}}{2\gamma}\omega_m \\  -\frac{M_{\rm S}}{2\gamma}\omega_m & Aq^2+B+\Pi(q,\omega_m) \end{pmatrix},
\end{equation}
and
\begin{equation}
    \mathcal{G}_0^{-1} = \left.\mathcal{G}^{-1}\right|_{m_y(x,\tau)=m_0}.\quad 
\end{equation}
Further
\begin{equation}
\begin{split}
    \Pi(q,\omega_m) = \frac{\lambda^2}{4} \frac{T}{L}\sum_{k,\varepsilon_n} \text{tr}_{2\times 2}&\Big[ \mathcal{G}_0(k,\varepsilon_n)\hat{\tau}_y\\&\mathcal{G}_0(k+q,\varepsilon_n+\omega_m)\hat{\tau}_y\Big]
\end{split}
\end{equation}
with $\omega_m$ and $\varepsilon_n$ being the bosonic and fermionic Matsubara frequencies respectively. The linear terms in $\delta m_y$ only contain $\delta m_y(q=0,\omega_m=0)$-contributions and can therefore safely be omitted when determining $m_0$.

From the partition function $Z = \int\mathcal{D}(\delta m_x,\delta m_y)\,e^{-S_{\rm eff}}$ one obtains the ground state energy
\begin{align}
    \frac{E_{\rm G}}{L} = &\left[B+\frac{\lambda^2}{4\pi v}\left(\log\frac{\lambda m_0}{2v\Lambda}-\frac{1}{2}\right)\right]m_0^2\\
    &+\left.\frac{1}{2}\frac{T}{L}\sum_{q,\omega_m}\log\text{det}_{2\times 2}\left(2\mathcal{D}_{q,\omega_m}^{-1}\right)\right|_{T\rightarrow 0} + \frac{E_0}{L}, \notag 
\end{align}
and the corresponding gap equation reads
\begin{equation}
    \frac{1}{L}\frac{\text{d}E_{\rm G}}{\text{d}(\Delta^2)} = \frac{1}{4\pi v}\log\frac{\Delta}{\tilde{\Delta}} + \chi = 0, \label{eq_gapeq}
\end{equation}
where $\Delta\equiv\lambda m_0$ and $\chi$ denotes the contribution due to the fluctuations
\begin{equation}
    \chi = \left.\frac{1}{2}\int\frac{\text{d}q\,\text{d}\omega}{(2\pi)^2}\,\frac{\frac{\text{d}}{\text{d}(\Delta^2)}\,\text{det}_{2\times 2}\mathcal{D}_{q,\omega}^{-1}}{\text{det}_{2\times 2}\mathcal{D}_{q,\omega}^{-1}}\right|_{T\rightarrow 0}. \label{eq_chi}
\end{equation}
Obviously, without fluctuations, the mean-field solution $\Delta=\tilde{\Delta}$ is recovered. In order to take the fluctuations into account, eq.\ (\ref{eq_chi}) and, thus, $\Pi(q,\omega)$ need to be evaluated. Following Ref.\ \onlinecite{vaks_collective_1962}, it can be seen that
\begin{equation}
    \Pi(q,\omega) = \frac{\lambda^2}{4\pi v} \left[\log\frac{\Delta}{2v\Lambda} + \frac{\sqrt{1+r^2}}{r}\text{Arsinh}(r)\right]\ ,
\end{equation}
with $r = \sqrt{v^2q^2+\omega^2}/2\Delta$ being the radial coordinate in the $\left(\frac{vq}{2\Delta},\frac{\omega}{2\Delta}\right)$-plane. It follows
\begin{widetext}
\begin{equation}
    \chi = \frac{\lambda^2}{4\pi v}\frac{4}{\pi}\int_0^\infty\text{d}r\,\int_0^{\pi/2}\text{d}\varphi\,\frac{\left(\tilde{A}\,\frac{\Delta^2}{\tilde{\Delta}^2}\,r^2\cos^2\varphi+1\right)\frac{\text{Arsinh}(r)}{\sqrt{1+r^2}}}{\left(\tilde{A}\,\frac{\Delta^2}{\tilde{\Delta}^2}\,r^2\cos^2\varphi+1\right)\left(\frac{4\pi vB}{\lambda^2}\,\tilde{A}\,\frac{\Delta^2}{\tilde{\Delta}^2}\,r^2\cos^2\varphi+\log\frac{\Delta}{\tilde{\Delta}}+\frac{\sqrt{1+r^2}}{r}\text{Arsinh}(r)\right)+\tilde{M}\,\frac{\Delta^2}{\tilde{\Delta}^2}\,r^2\sin^2\varphi} \label{eq_chiintegral}
\end{equation}
\end{widetext}
with the dimensionless parameters $\tilde{A} = \frac{4\tilde{\Delta}^2}{v^2B}A$ and $\tilde{M} = \frac{4\pi v \tilde{\Delta}^2}{\lambda^2B}\frac{M_{\rm S}^2}{\gamma^2}$. The solution to the gap equation relative to the mean-field solution $\Delta/\tilde{\Delta}$, which corresponds to the minimum of the ground state energy, thus depends on the values of $\tilde{A}$, $\tilde{M}$ and the `BCS parameter' $\frac{4\pi v B}{\lambda^2}$. If no solution to eq.\ (\ref{eq_gapeq}) exists, i.e. if $\chi$, which is always positive, is too large to be compensated by the logarithm in (\ref{eq_gapeq}), the minimum of the ground state energy is shifted back to $\Delta=0$ due to the fluctuations. 

Note that for $A=M_{\rm S}/\gamma=0$ the integral (\ref{eq_chiintegral}) is logarithmically UV-divergent, complying with the results in BCS theory in Ref.\ \onlinecite{kos_gaussian_2004}. This divergence is remedied as soon as either $A$ or $M_{\rm S}/\gamma$ take on a finite value, in accordance with our conjecture that either of these parameters allow the fluctuations to be controlled. The numerical solutions to the gap equation for different combinations of the dimensionless parameters can be seen in Fig.\ \ref{fig:numsolution_gapeq}. We find that indeed for small values of $\tilde{A}$ and $\tilde{M}$ the fluctuations do not allow for any notion of spontaneous symmetry breaking and mean-field theory fails completely, while above a certain threshold (even if either of the parameters vanishes), the minima of the ground state energy persist and the only effect is a lowering of the Majorana mass gap $\Delta$ to values as low as $\Delta/\tilde{\Delta}\sim 0.6$.

This threshold can be interpreted as the transition region, separating the two regimes (I) without broken symmetry, with short magnetic coherence length and free massless Majorana modes and (II) with broken magnetic symmetry and spin canting, leading to a finite ground state expectation value $\braket{\lambda m_y} = \Delta$, establishing a finite Majorana mass gap and a magnetic coherence length which is exponentially large at low temperatures.

In Fig.\ \ref{fig:numsolution_gapeq}, the quantum phase transition between (I) and (II) seems to be of first order with the order parameter discontinuously jumping to zero at the boundary. In Ref.\ \onlinecite{rahmani_phase_2015}, an apparently related phase transition in Majorana chains with minimally nonlocal interactions has been found to be described by the tricritial Ising conformal field theory with central charge $c=7/10$. Most likely, the Gaussian approximation employed here cannot be trusted in the vicinity of the phase transition, but only provides evidence for the existence of the two distinct phases. In addition, it offers an estimate on the parameter regime where the transition takes place.

An additional insight into the problem is provided by analyzing the relation between the Higgs frequency of the order parameter and the fermionic gap. It is well known that in the case of phonon-induced superconductivity, the Higgs mass, which is the frequency of the Higgs mode at $q=0$, is given by $\omega(q=0)=2\tilde{\Delta}$ and thus lies exactly on the edge of the quasiparticle continuum\cite{kleinert_collective_1978,varma_higgs_2002}. In contrast, in Peierls systems, $\omega(q=0)\ll \tilde{\Delta}$. This result has been obtained in Refs.\ \onlinecite{lee_conductivity_1974} and \onlinecite{brazovskii_dynamics_1976} and was also discussed in detail in Ref.\ \onlinecite{gogolin_bosonization_2004}. In our case, the Higgs mass is determined by $\text{det}\,\mathcal{D}^{-1}_{q=0,i\omega}=0$. For $\Delta=\tilde{\Delta}$, $M_{\rm S}/\gamma=0$, it follows to be $\omega(q=0)=2\tilde{\Delta}$. A non-zero value of $M_{\rm S}/\gamma$ now reduces the Higgs mass, as illustrated in Fig.\ \ref{fig:higgsgap}, in accordance with the above cited works on Peierls systems. 
This departure from the continuum leads to the Higgs mode being underdamped. Increasing the value of $A$ additionally shifts the spectral weight of the fluctuations to lower energies.


\section{Discontinuity in the current-phase relation}

\begin{figure*}[ht]
    \centering
    \subfloat[\label{fig:doublewelleps}]{
    \includegraphics[width=0.45\textwidth]{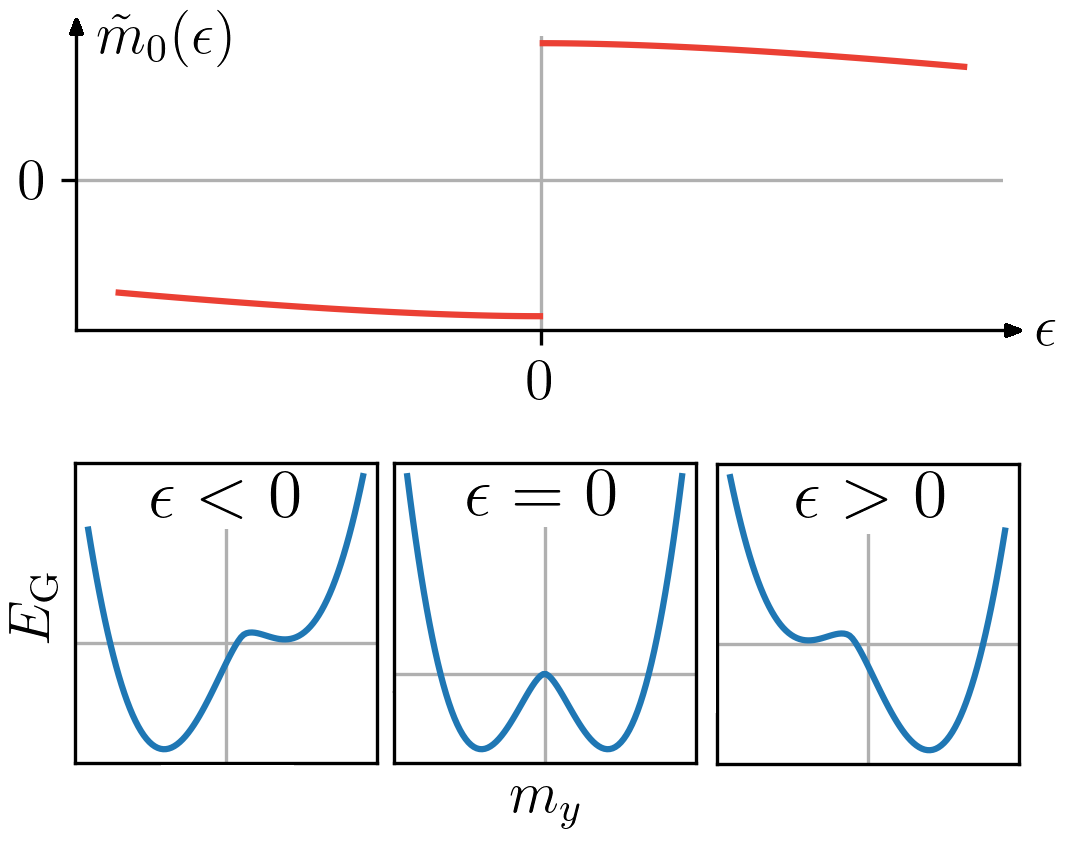}}
    \hspace{0.5cm}
    \subfloat[\label{fig:currentphasejump}]{
    \includegraphics[width=0.45\textwidth]{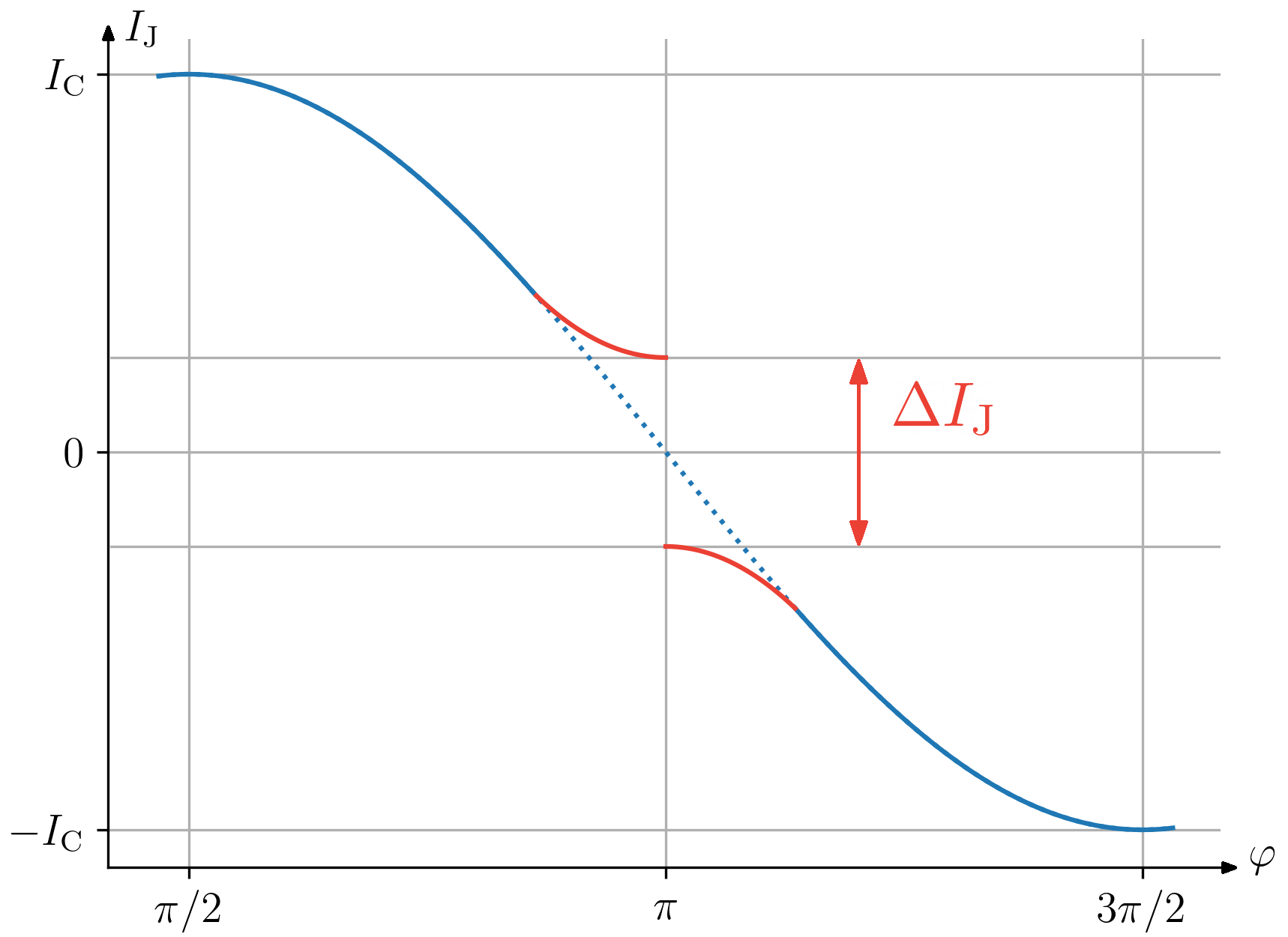}}
    \caption{a) Allowing for the phase difference $\varphi$ between the superconductors to deviate from $\pi$ by $\epsilon=\pi-\varphi$ lifts the ground state degeneracy. The mean-field value of $m_y$ is positive (negative) for positive (negative) $\epsilon$ with a first order transition at $\epsilon=0$. b) At mean-field level, this leads to a discontinuity of magnitude $\Delta I_{\rm J}=8eBL\tilde{m}_0$ in the current-phase relation, which could be probed experimentally.}
    \label{fig:jump}
\end{figure*}

Until now we have assumed the phase difference between the superconductors to be fixed to $\varphi=\pi$. In Appendix \ref{app_deriving-effham}, it is shown that allowing $\epsilon=\pi-\varphi$ to take on a small, non-zero value leads to the following hybridization between the Majorana modes:
\begin{align}
    H_{\rm hybr.} = -i\lambda\int\text{d}x\,\left(m_y+\frac{v_{\rm F}}{2MW}\epsilon\right)\gamma_{\rm L}\gamma_{\rm R}.
\end{align}
The ground state energy then is given by
\begin{align}
        &\frac{E_{\rm G}}{L} = \frac{E_0}{L}+Bm_y^2\,+\\&+\frac{\lambda^2}{4\pi v}\left(\log\left|\frac{\lambda \left(m_y+\frac{v_{\rm F}}{2MW}\epsilon\right)}{2v\Lambda}\right|-\frac{1}{2}\right)\left(m_y+\frac{v_{\rm F}}{2MW}\epsilon\right)^2 .\notag
\end{align}
As sketched in Fig.~\ref{fig:doublewelleps}, a deviation of $\varphi$ from $\pi$ thus lifts the degeneracy of the ground state and, in the mean-field approximation, $m_y$ takes on a value $\tilde{m}_0(\epsilon> 0)> 0$ and vice versa. The transition at $\epsilon=0$ is of first order.

The Josephson current is given by $I_{\rm J} = 2e\,\partial F/\partial\varphi$, where $F$ is the free energy. At zero temperature, the current carried by the Majorana modes near $\varphi=\pi$ consequently reads
\begin{align}
    \frac{I_{\rm J}^{\rm Maj}}{L} = -\frac{2e}{L}\frac{\partial E_G}{\partial \epsilon} = 4eB\tilde{m}_0(\epsilon).
\end{align}
In particular, it exhibits a discontinuity at $\epsilon = 0$. In the parameter regime (II), this discontinuous jump persists when taking Gaussian fluctuations of $m_y$ around the mean-field value into account. As it is an IR effect, the discontinuity will furthermore not be compensated by considering the higher-energy scattering states in addition to the Majorana modes. A sketch of the expected current-phase relation is provided in Fig.~\ref{fig:currentphasejump}. 

Measuring the Josephson current in the phase-biased junction and examining it for a discontinuous jump thus provides a possibility to experimentally confirm the Ising-like properties we predict.

In order to estimate the magnitude of the jump compared to the critical Josephson current, let us for simplicity assume that $K\approx 1$, where $K$ is the parameter defined in Appendix \ref{app_deriving-effham}. This is the case if for example $\sqrt{\mu^2-M^2}W/v_{\rm F}\approx n\cdot2\pi, n\in \mathbb{N}$. The ratio between the jump $\Delta I_{\rm J} = 8eBL\tilde{m}_0(\epsilon=0)$ and the characteristic current scale $I_0 = eL\Delta_0^2/v$ (as defined in Ref.~\onlinecite{backens_current--phase_2021}) is then given by
\begin{align}
    \frac{\Delta I_{\rm J}}{I_0} = \frac{4}{\pi} \frac{\lambda}{\Delta_0}\frac{4\pi vB}{\lambda^2}e^{-4\pi vB/\lambda^2},
\end{align}
where $\lambda/\Delta_0 = MW/v_{\rm F}$ and we took the cut-off to be $v\Lambda = \Delta_0$. Now, suppose $4\pi vB/\lambda^2=10$, as we did above. The Fermi velocity of the topological insulator Bi$_2$Se$_3$ has been experimentally determined to be $\hbar v_{\rm F} \approx 0.3\,{\rm eV\,nm}$ with a Fermi energy of $\mu \approx 0.3\,{\rm eV}$.\cite{zhang_topological_2008,xia_electrons_2008} Assuming the junction to have a width of $W\sim 1\,{\rm \mu m}$, it follows that a value of $\Delta I_{\rm J}/I_0 \sim 0.1$ could be achieved with a magnetization energy of $M\approx 0.1\,{\rm eV}$, which seems experimentally feasible. 

\section{Magnetic solitons \label{sec_solitons}}
In the parameter regime (II) of Fig.~\ref{fig:numsolutions} we argued, in analogy with the Ising model, the magnetic coherence length to be given by $\xi\sim b\,e^{E_{\rm DW}/T}$ with $E_{\rm DW}$ and $b$ the energy and width of a domain wall between regions with $\braket{m_y}=\pm m_0$ respectively. Such domain walls, which are responsible for the lack of long-range order at $T>0$, correspond in mean-field theory to saddle points of the effective action (\ref{eq_trlogeffaction}) with non-constant $m_y$. For static but spatially varying configurations $m_y=m_0(x)$ which extremize the effective action, it holds
\begin{equation}
\begin{split}
    (-A\partial_x^2&+B)\,m_0(x)\\ &= \frac{i\lambda}{4}\sum_ju_j(x)v_j^\ast(x)\tanh(E_j/2T)
\end{split}
\end{equation}
with $(u_j(x),v_j(x))$ being the solutions to the BdG equation in $(\gamma_{\rm R},\gamma_{\rm L})$-space with eigenenergies $E_j$
\begin{equation}
\begin{split}
\begin{pmatrix}-iv\partial_x & i\lambda m_0(x) \\ -i\lambda m_0(x) & iv\partial_x\end{pmatrix}\begin{pmatrix} u_j(x) \\ v_j(x) \end{pmatrix}= E_j \begin{pmatrix} u_j(x) \\ v_j(x) \end{pmatrix}.
\end{split}
\end{equation}
A single domain wall, or soliton, in the system is then given by the configuration $m_0(x)$ self-consistently solving these two equations and asymptotically approaching the mean-field solutions $m_0(x\rightarrow\pm\infty) = \pm \tilde{m}_0$ or $m_0(x\rightarrow\pm\infty) = \mp \tilde{m}_0$ with a single switch of the sign at some value of $x$, which we take to be 0 without loss of generality.

In this case the BdG solutions generally consist of a continuous spectrum for energies $|E_j|>\lambda \tilde{m}_0$, one zero-energy Majorana bound state (MBS) as well as other Andreev bound states (ABS) with discrete non-zero in-gap energies localized near $x=0$, where the number of ABS is dependent on the width of the soliton, while the MBS is always present at a zero-crossing \cite{jackiw_solitons_1976,grosfeld_observing_2011}.

For the case $A=0$, in the present model with the fluctuations thus dampened by a sufficiently large $\tilde{M}$, there is an extensive literature \cite{zakharov_integrability_1980,brazovskii_self-localized_1980,takayama_continuum_1980,takahashi_self-consistent_2013,takahashi_exhaustive_2016} on the exact one-soliton and multi-soliton solutions to this and generalized problems. The solutions are obtained by employing methods of inverse scattering theory. The one-soliton solution is shown to read
\begin{equation}
    m_0(x) = \pm\tilde{m}_0\tanh(x/b)
\end{equation}
with the width of the domain wall given by $b=\frac{v}{\lambda \tilde{m}_0}$. This soliton only carries a single bound state, namely a  MBS. The energy $E_{\rm DW}$ is given by the difference between the mean-field energy in presence of a soliton $E_{\rm MF}[m_0(x)]$ and the ground-state energy $E_{\rm G}=E_{\rm MF}[\tilde{m}_0]$ and follows to be \cite{takayama_continuum_1980}
\begin{equation}
    E_{\rm DW} = E_{\rm MF}[m_0(x)]-E_{\rm MF}[\tilde{m}_0] = \frac{\lambda \tilde{m}_0}{2\pi},
\end{equation}
wherein the zero-energy state also contributes by lowering the continuum density of states through its appearance.

It is to be expected that a small, non-zero value of $A$ will in a first approximation only alter the length scale of the transition, making it wider and at some point leading to additional bound states to arise, while the overall shape of the soliton is preserved. 

Finding a uniformly moving soliton-solution is non-trivial and cannot be achieved by a simple `boost' of the stationary solution, as only the fermionic part of the action is Lorentz invariant, while the magnetic action possesses Galilean invariance in the following sense: if $(m_x(x,t),m_y(x,t))^T$ is a solution to the (real-time) equations of motion of the free magnetization field, so is
\begin{equation}
\begin{pmatrix}
\cos\phi_u(x,t) & -\sin\phi_u(x,t) \\ \sin\phi_u(x,t) & \cos\phi_u(x,t)
\end{pmatrix}\begin{pmatrix}
m_x(x-ut) \\ m_y(x-ut)
\end{pmatrix},
\end{equation}
where $\phi_u(x,t)=\frac{uM_{\rm S}}{4\gamma A}\left(x-\frac{u}{2}t\right)$ (this is analogous to the Galilean invariance of the 1D nonlinear Schr\"odinger equation). The coupling between $m_x$ and $m_y$, present due to $M_{\rm S}/\gamma$ if either of the magnetization field components is time dependent, therefore necessitates a moving soliton to include rotation of the magnetization around the $z$-axis.

However, as is the case deep in the ordered phase of the transverse-field Ising model, the dynamical soliton mass can be expected to be very large and the inclusion of only stationary solutions to the statistical argument thus to provide a good approximation.

We leave further analysis of the cases with non-zero $A$ and non-stationary solitons for possible future work.

\section{Summary and Conclusions}
In this paper we have studied the coupling of the dynamic magnetization of a ferromagnet deposited in a topological Josephson junction to the low-energy Majorana modes in the surface states of such a structure. We saw that the component of the magnetization field perpendicular to the junction acts as a Majorana mass term, hybridizing the right- and left-moving modes. In a mean-field treatment, this component of the magnetization field takes on a finite ground state value, signifying an instability of the magnetic easy axis and an ensuing opening of a Majorana mass gap.  This is the Majorana analog to the commensurate case of the Peierls instability in the 1D Fröhlich model. A fluctuation analysis of the associated gap equation at zero temperature provides evidence for the existence of a cross-over from a parameter regime where the instability does not persist and the interaction between magnetic and fermionic degrees of freedom is irrelevant, to a regime where the mean-field value is diminished but stabilized due to either a large spatial or temporal stiffness of the magnetization dynamics. A possible experimental signature of this broken symmetry is a discontinuous jump in the corresponding current-phase relation. In analogy to the Ising model at low but finite temperatures, we argued that there are exponentially large stretches of ordered phase, which are separated by domain walls, each carrying a MBS. The energy and width of these domain walls allow an estimate of the magnetic coherence length. We furthermore provided the corresponding self-consistency problem in the case of stationary solitons, for which the solution is known if the exchange coupling can be neglected.

These considerations are not exclusive to the suggested system but generalizable to other models of coupled fermions and bosons in one spatial dimension. For a more thorough picture on the existence and nature of the phase transition, however, more sophisticated methods than the here employed fluctuation analysis will be needed. 

\section*{Acknowledgments}

We thank Liliana Arrachea, Andrey V. Chubukov, Yuriy Makhlin,  Alexander Mirlin, and Michael Rampp for fruitful discussions. 
This research was supported by the DFG under the grant No.~SH81/7-1. J.\ S.\ acknowledges support by the Deutsche Forschungsgemeinschaft (German Research Foundation) Project ER 463/14-1. E.\ B.\ was supported by the Israel Science Foundation Quantum Science and Technology grant no. 2074/19, and by the CRC 183 of the Deutsche Forschungsgemeinschaft (Project C02).

\bibliography{references.bib}

\begin{thebibliography}{40}
\expandafter\ifx\csname natexlab\endcsname\relax\def\natexlab#1{#1}\fi
\expandafter\ifx\csname bibnamefont\endcsname\relax
  \def\bibnamefont#1{#1}\fi
\expandafter\ifx\csname bibfnamefont\endcsname\relax
  \def\bibfnamefont#1{#1}\fi
\expandafter\ifx\csname citenamefont\endcsname\relax
  \def\citenamefont#1{#1}\fi
\expandafter\ifx\csname url\endcsname\relax
  \def\url#1{\texttt{#1}}\fi
\expandafter\ifx\csname urlprefix\endcsname\relax\def\urlprefix{URL }\fi
\providecommand{\bibinfo}[2]{#2}
\providecommand{\eprint}[2][]{\url{#2}}

\bibitem[{\citenamefont{Fu and Kane}(2008)}]{fu_superconducting_2008}
\bibinfo{author}{\bibfnamefont{L.}~\bibnamefont{Fu}} \bibnamefont{and}
  \bibinfo{author}{\bibfnamefont{C.~L.} \bibnamefont{Kane}},
  \bibinfo{journal}{Physical Review Letters} \textbf{\bibinfo{volume}{100}},
  \bibinfo{pages}{096407} (\bibinfo{year}{2008}),
  \urlprefix\url{https://link.aps.org/doi/10.1103/PhysRevLett.100.096407}.

\bibitem[{\citenamefont{Fu and Kane}(2009)}]{fu_probing_2009}
\bibinfo{author}{\bibfnamefont{L.}~\bibnamefont{Fu}} \bibnamefont{and}
  \bibinfo{author}{\bibfnamefont{C.~L.} \bibnamefont{Kane}},
  \bibinfo{journal}{Physical Review Letters} \textbf{\bibinfo{volume}{102}},
  \bibinfo{pages}{216403} (\bibinfo{year}{2009}), ISSN
  \bibinfo{issn}{0031-9007, 1079-7114},
  \urlprefix\url{https://journals.aps.org/prl/abstract/10.1103/PhysRevLett.102.216403}.

\bibitem[{\citenamefont{Tanaka et~al.}(2009)\citenamefont{Tanaka, Yokoyama, and
  Nagaosa}}]{tanaka_manipulation_2009}
\bibinfo{author}{\bibfnamefont{Y.}~\bibnamefont{Tanaka}},
  \bibinfo{author}{\bibfnamefont{T.}~\bibnamefont{Yokoyama}}, \bibnamefont{and}
  \bibinfo{author}{\bibfnamefont{N.}~\bibnamefont{Nagaosa}},
  \bibinfo{journal}{Physical Review Letters} \textbf{\bibinfo{volume}{103}},
  \bibinfo{pages}{107002} (\bibinfo{year}{2009}), ISSN
  \bibinfo{issn}{0031-9007, 1079-7114},
  \urlprefix\url{https://link.aps.org/doi/10.1103/PhysRevLett.103.107002}.

\bibitem[{\citenamefont{Potter and Fu}(2013)}]{potter_anomalous_2013}
\bibinfo{author}{\bibfnamefont{A.~C.} \bibnamefont{Potter}} \bibnamefont{and}
  \bibinfo{author}{\bibfnamefont{L.}~\bibnamefont{Fu}},
  \bibinfo{journal}{Physical Review B} \textbf{\bibinfo{volume}{88}},
  \bibinfo{pages}{121109} (\bibinfo{year}{2013}), ISSN
  \bibinfo{issn}{1098-0121, 1550-235X},
  \urlprefix\url{https://link.aps.org/doi/10.1103/PhysRevB.88.121109}.

\bibitem[{\citenamefont{Hegde et~al.}(2020)\citenamefont{Hegde, Yue, Wang,
  Huemiller, Van~Harlingen, and Vishveshwara}}]{hegde_topological_2020}
\bibinfo{author}{\bibfnamefont{S.~S.} \bibnamefont{Hegde}},
  \bibinfo{author}{\bibfnamefont{G.}~\bibnamefont{Yue}},
  \bibinfo{author}{\bibfnamefont{Y.}~\bibnamefont{Wang}},
  \bibinfo{author}{\bibfnamefont{E.}~\bibnamefont{Huemiller}},
  \bibinfo{author}{\bibfnamefont{D.~J.} \bibnamefont{Van~Harlingen}},
  \bibnamefont{and}
  \bibinfo{author}{\bibfnamefont{S.}~\bibnamefont{Vishveshwara}},
  \bibinfo{journal}{Annals of Physics} \textbf{\bibinfo{volume}{423}},
  \bibinfo{pages}{168326} (\bibinfo{year}{2020}), ISSN
  \bibinfo{issn}{00034916},
  \urlprefix\url{https://www.sciencedirect.com/science/article/abs/pii/S0003491620302608}.

\bibitem[{\citenamefont{Backens and
  Shnirman}(2021)}]{backens_current--phase_2021}
\bibinfo{author}{\bibfnamefont{S.}~\bibnamefont{Backens}} \bibnamefont{and}
  \bibinfo{author}{\bibfnamefont{A.}~\bibnamefont{Shnirman}},
  \bibinfo{journal}{Physical Review B} \textbf{\bibinfo{volume}{103}},
  \bibinfo{pages}{115423} (\bibinfo{year}{2021}),
  \urlprefix\url{https://link.aps.org/doi/10.1103/PhysRevB.103.115423}.

\bibitem[{\citenamefont{Zyuzin et~al.}(2016)\citenamefont{Zyuzin, Alidoust, and
  Loss}}]{zyuzin_josephson_2016}
\bibinfo{author}{\bibfnamefont{A.}~\bibnamefont{Zyuzin}},
  \bibinfo{author}{\bibfnamefont{M.}~\bibnamefont{Alidoust}}, \bibnamefont{and}
  \bibinfo{author}{\bibfnamefont{D.}~\bibnamefont{Loss}},
  \bibinfo{journal}{Physical Review B} \textbf{\bibinfo{volume}{93}},
  \bibinfo{pages}{214502} (\bibinfo{year}{2016}), ISSN
  \bibinfo{issn}{2469-9950, 2469-9969},
  \urlprefix\url{https://journals.aps.org/prb/abstract/10.1103/PhysRevB.93.214502}.

\bibitem[{\citenamefont{Bobkova et~al.}(2016)\citenamefont{Bobkova, Bobkov,
  Zyuzin, and Alidoust}}]{bobkova_magnetoelectrics_2016}
\bibinfo{author}{\bibfnamefont{I.~V.} \bibnamefont{Bobkova}},
  \bibinfo{author}{\bibfnamefont{A.~M.} \bibnamefont{Bobkov}},
  \bibinfo{author}{\bibfnamefont{A.~A.} \bibnamefont{Zyuzin}},
  \bibnamefont{and} \bibinfo{author}{\bibfnamefont{M.}~\bibnamefont{Alidoust}},
  \bibinfo{journal}{Physical Review B} \textbf{\bibinfo{volume}{94}},
  \bibinfo{pages}{134506} (\bibinfo{year}{2016}), ISSN
  \bibinfo{issn}{2469-9950, 2469-9969},
  \urlprefix\url{https://journals.aps.org/prb/abstract/10.1103/PhysRevB.94.134506}.

\bibitem[{\citenamefont{Amundsen et~al.}(2018)\citenamefont{Amundsen, Hugdal,
  Sudbø, and Linder}}]{amundsen_vortex_2018}
\bibinfo{author}{\bibfnamefont{M.}~\bibnamefont{Amundsen}},
  \bibinfo{author}{\bibfnamefont{H.~G.} \bibnamefont{Hugdal}},
  \bibinfo{author}{\bibfnamefont{A.}~\bibnamefont{Sudbø}}, \bibnamefont{and}
  \bibinfo{author}{\bibfnamefont{J.}~\bibnamefont{Linder}},
  \bibinfo{journal}{Physical Review B} \textbf{\bibinfo{volume}{98}},
  \bibinfo{pages}{144505} (\bibinfo{year}{2018}), ISSN
  \bibinfo{issn}{2469-9950, 2469-9969},
  \urlprefix\url{https://link.aps.org/doi/10.1103/PhysRevB.98.144505}.

\bibitem[{\citenamefont{Landau and Lifshitz}(1935)}]{landau_theory_1935}
\bibinfo{author}{\bibfnamefont{L.}~\bibnamefont{Landau}} \bibnamefont{and}
  \bibinfo{author}{\bibfnamefont{E.}~\bibnamefont{Lifshitz}},
  \bibinfo{journal}{Phys. Z. Sowjetunion} \textbf{\bibinfo{volume}{8}},
  \bibinfo{pages}{153} (\bibinfo{year}{1935}).

\bibitem[{\citenamefont{Gilbert}(1956)}]{gilbert_formulations_1956}
\bibinfo{author}{\bibfnamefont{T.}~\bibnamefont{Gilbert}},
  \bibinfo{journal}{Ph. D. dissertation}  (\bibinfo{year}{1956}).

\bibitem[{\citenamefont{Gilbert}(2004)}]{gilbert_phenomenological_2004}
\bibinfo{author}{\bibfnamefont{T.}~\bibnamefont{Gilbert}},
  \bibinfo{journal}{IEEE Transactions on Magnetics}
  \textbf{\bibinfo{volume}{40}}, \bibinfo{pages}{3443} (\bibinfo{year}{2004}),
  ISSN \bibinfo{issn}{0018-9464},
  \urlprefix\url{http://ieeexplore.ieee.org/document/1353448/}.

\bibitem[{\citenamefont{Konschelle and
  Buzdin}(2009)}]{konschelle_magnetic_2009}
\bibinfo{author}{\bibfnamefont{F.}~\bibnamefont{Konschelle}} \bibnamefont{and}
  \bibinfo{author}{\bibfnamefont{A.}~\bibnamefont{Buzdin}},
  \bibinfo{journal}{Physical Review Letters} \textbf{\bibinfo{volume}{102}},
  \bibinfo{pages}{017001} (\bibinfo{year}{2009}), ISSN
  \bibinfo{issn}{0031-9007, 1079-7114},
  \urlprefix\url{https://link.aps.org/doi/10.1103/PhysRevLett.102.017001}.

\bibitem[{\citenamefont{Nashaat et~al.}(2019)\citenamefont{Nashaat, Bobkova,
  Bobkov, Shukrinov, Rahmonov, and Sengupta}}]{nashaat_electrical_2019}
\bibinfo{author}{\bibfnamefont{M.}~\bibnamefont{Nashaat}},
  \bibinfo{author}{\bibfnamefont{I.~V.} \bibnamefont{Bobkova}},
  \bibinfo{author}{\bibfnamefont{A.~M.} \bibnamefont{Bobkov}},
  \bibinfo{author}{\bibfnamefont{Y.~M.} \bibnamefont{Shukrinov}},
  \bibinfo{author}{\bibfnamefont{I.~R.} \bibnamefont{Rahmonov}},
  \bibnamefont{and} \bibinfo{author}{\bibfnamefont{K.}~\bibnamefont{Sengupta}},
  \bibinfo{journal}{Physical Review B} \textbf{\bibinfo{volume}{100}},
  \bibinfo{pages}{054506} (\bibinfo{year}{2019}), ISSN
  \bibinfo{issn}{2469-9950, 2469-9969},
  \urlprefix\url{https://link.aps.org/doi/10.1103/PhysRevB.100.054506}.

\bibitem[{\citenamefont{Fröhlich}(1954)}]{frohlich_theory_1954}
\bibinfo{author}{\bibfnamefont{H.}~\bibnamefont{Fröhlich}},
  \bibinfo{journal}{Proceedings of the Royal Society of London. Series A.
  Mathematical and Physical Sciences} \textbf{\bibinfo{volume}{223}},
  \bibinfo{pages}{296} (\bibinfo{year}{1954}), ISSN \bibinfo{issn}{0080-4630,
  2053-9169},
  \urlprefix\url{https://royalsocietypublishing.org/doi/10.1098/rspa.1954.0116}.

\bibitem[{\citenamefont{Ruiz et~al.}(2022)\citenamefont{Ruiz, Rampp, Aligia,
  Schmalian, and Arrachea}}]{ruiz_josephson_2022}
\bibinfo{author}{\bibfnamefont{G.~F.~R.} \bibnamefont{Ruiz}},
  \bibinfo{author}{\bibfnamefont{M.~A.} \bibnamefont{Rampp}},
  \bibinfo{author}{\bibfnamefont{A.~A.} \bibnamefont{Aligia}},
  \bibinfo{author}{\bibfnamefont{J.}~\bibnamefont{Schmalian}},
  \bibnamefont{and} \bibinfo{author}{\bibfnamefont{L.}~\bibnamefont{Arrachea}},
  \bibinfo{journal}{Physical Review B} \textbf{\bibinfo{volume}{106}},
  \bibinfo{pages}{195415} (\bibinfo{year}{2022}), ISSN
  \bibinfo{issn}{2469-9950, 2469-9969},
  \urlprefix\url{https://link.aps.org/doi/10.1103/PhysRevB.106.195415}.

\bibitem[{\citenamefont{Kos et~al.}(2004)\citenamefont{Kos, Millis, and
  Larkin}}]{kos_gaussian_2004}
\bibinfo{author}{\bibfnamefont{S.}~\bibnamefont{Kos}},
  \bibinfo{author}{\bibfnamefont{A.~J.} \bibnamefont{Millis}},
  \bibnamefont{and} \bibinfo{author}{\bibfnamefont{A.~I.}
  \bibnamefont{Larkin}}, \bibinfo{journal}{Physical Review B}
  \textbf{\bibinfo{volume}{70}}, \bibinfo{pages}{214531}
  (\bibinfo{year}{2004}), ISSN \bibinfo{issn}{1098-0121, 1550-235X},
  \urlprefix\url{https://link.aps.org/doi/10.1103/PhysRevB.70.214531}.

\bibitem[{\citenamefont{Fischer et~al.}(2018)\citenamefont{Fischer, Hecker,
  Hoyer, and Schmalian}}]{fischer_short-distance_2018}
\bibinfo{author}{\bibfnamefont{S.}~\bibnamefont{Fischer}},
  \bibinfo{author}{\bibfnamefont{M.}~\bibnamefont{Hecker}},
  \bibinfo{author}{\bibfnamefont{M.}~\bibnamefont{Hoyer}}, \bibnamefont{and}
  \bibinfo{author}{\bibfnamefont{J.}~\bibnamefont{Schmalian}},
  \bibinfo{journal}{Physical Review B} \textbf{\bibinfo{volume}{97}},
  \bibinfo{pages}{054510} (\bibinfo{year}{2018}), ISSN
  \bibinfo{issn}{2469-9950, 2469-9969},
  \urlprefix\url{https://journals.aps.org/prb/abstract/10.1103/PhysRevB.97.054510}.

\bibitem[{\citenamefont{Alicea}(2012)}]{alicea_new_2012}
\bibinfo{author}{\bibfnamefont{J.}~\bibnamefont{Alicea}},
  \bibinfo{journal}{Reports on Progress in Physics}
  \textbf{\bibinfo{volume}{75}}, \bibinfo{pages}{076501}
  (\bibinfo{year}{2012}), ISSN \bibinfo{issn}{0034-4885, 1361-6633},
  \urlprefix\url{https://iopscience.iop.org/article/10.1088/0034-4885/75/7/076501}.

\bibitem[{\citenamefont{Gongyo et~al.}(2016)\citenamefont{Gongyo, Kikuchi,
  Hyodo, and Kunihiro}}]{gongyo_effective_2016}
\bibinfo{author}{\bibfnamefont{S.}~\bibnamefont{Gongyo}},
  \bibinfo{author}{\bibfnamefont{Y.}~\bibnamefont{Kikuchi}},
  \bibinfo{author}{\bibfnamefont{T.}~\bibnamefont{Hyodo}}, \bibnamefont{and}
  \bibinfo{author}{\bibfnamefont{T.}~\bibnamefont{Kunihiro}},
  \bibinfo{journal}{Progress of Theoretical and Experimental Physics}
  \textbf{\bibinfo{volume}{2016}}, \bibinfo{pages}{083B01}
  (\bibinfo{year}{2016}), ISSN \bibinfo{issn}{2050-3911},
  \urlprefix\url{https://academic.oup.com/ptep/article/2016/8/083B01/2594880}.

\bibitem[{\citenamefont{Rahmani et~al.}(2015)\citenamefont{Rahmani, Zhu, Franz,
  and Affleck}}]{rahmani_phase_2015}
\bibinfo{author}{\bibfnamefont{A.}~\bibnamefont{Rahmani}},
  \bibinfo{author}{\bibfnamefont{X.}~\bibnamefont{Zhu}},
  \bibinfo{author}{\bibfnamefont{M.}~\bibnamefont{Franz}}, \bibnamefont{and}
  \bibinfo{author}{\bibfnamefont{I.}~\bibnamefont{Affleck}},
  \bibinfo{journal}{Physical Review B} \textbf{\bibinfo{volume}{92}},
  \bibinfo{pages}{235123} (\bibinfo{year}{2015}), ISSN
  \bibinfo{issn}{1098-0121, 1550-235X},
  \urlprefix\url{https://link.aps.org/doi/10.1103/PhysRevB.92.235123}.

\bibitem[{\citenamefont{McKenzie}(1995)}]{mckenzie_microscopic_1995}
\bibinfo{author}{\bibfnamefont{R.~H.} \bibnamefont{McKenzie}},
  \bibinfo{journal}{Physical Review B} \textbf{\bibinfo{volume}{52}},
  \bibinfo{pages}{16428} (\bibinfo{year}{1995}), ISSN \bibinfo{issn}{0163-1829,
  1095-3795},
  \urlprefix\url{https://link.aps.org/doi/10.1103/PhysRevB.52.16428}.

\bibitem[{\citenamefont{Peierls}(1936)}]{peierls_isings_1936}
\bibinfo{author}{\bibfnamefont{R.}~\bibnamefont{Peierls}},
  \bibinfo{journal}{Mathematical Proceedings of the Cambridge Philosophical
  Society} \textbf{\bibinfo{volume}{32}}, \bibinfo{pages}{477}
  (\bibinfo{year}{1936}), ISSN \bibinfo{issn}{0305-0041, 1469-8064},
  \urlprefix\url{https://www.cambridge.org/core/product/identifier/S0305004100019174/type/journal_article}.

\bibitem[{\citenamefont{Griffiths}(1964)}]{griffiths_peierls_1964}
\bibinfo{author}{\bibfnamefont{R.~B.} \bibnamefont{Griffiths}},
  \bibinfo{journal}{Physical Review} \textbf{\bibinfo{volume}{136}},
  \bibinfo{pages}{A437} (\bibinfo{year}{1964}), ISSN \bibinfo{issn}{0031-899X},
  \urlprefix\url{https://link.aps.org/doi/10.1103/PhysRev.136.A437}.

\bibitem[{\citenamefont{Lee et~al.}(1974)\citenamefont{Lee, Rice, and
  Anderson}}]{lee_conductivity_1974}
\bibinfo{author}{\bibfnamefont{P.~A.} \bibnamefont{Lee}},
  \bibinfo{author}{\bibfnamefont{T.}~\bibnamefont{Rice}}, \bibnamefont{and}
  \bibinfo{author}{\bibfnamefont{P.~W.} \bibnamefont{Anderson}},
  \bibinfo{journal}{Solid State Communications} \textbf{\bibinfo{volume}{14}},
  \bibinfo{pages}{703} (\bibinfo{year}{1974}).

\bibitem[{\citenamefont{Brazovskiĭ and
  Dzyaloshinskiǐ}(1976)}]{brazovskii_dynamics_1976}
\bibinfo{author}{\bibfnamefont{S.~A.} \bibnamefont{Brazovskiĭ}}
  \bibnamefont{and} \bibinfo{author}{\bibfnamefont{I.~E.}
  \bibnamefont{Dzyaloshinskiǐ}}, \bibinfo{journal}{Zh. Eksp. Teor. Fiz.}
  \textbf{\bibinfo{volume}{71}}, \bibinfo{pages}{2338} (\bibinfo{year}{1976}).

\bibitem[{\citenamefont{Gogolin et~al.}(2004)\citenamefont{Gogolin, Nersesyan,
  and Tsvelik}}]{gogolin_bosonization_2004}
\bibinfo{author}{\bibfnamefont{A.~O.} \bibnamefont{Gogolin}},
  \bibinfo{author}{\bibfnamefont{A.~A.} \bibnamefont{Nersesyan}},
  \bibnamefont{and} \bibinfo{author}{\bibfnamefont{A.~M.}
  \bibnamefont{Tsvelik}}, \emph{\bibinfo{title}{Bosonization and strongly
  correlated systems}} (\bibinfo{publisher}{Cambridge University Press},
  \bibinfo{address}{Cambridge}, \bibinfo{year}{2004}), ISBN
  \bibinfo{isbn}{978-0-521-61719-2 978-0-521-59031-0}.

\bibitem[{\citenamefont{Reich et~al.}(2023)\citenamefont{Reich, Schmalian, and
  Shnirman}}]{reich_unpublished_2023}
\bibinfo{author}{\bibfnamefont{A.}~\bibnamefont{Reich}},
  \bibinfo{author}{\bibfnamefont{J.}~\bibnamefont{Schmalian}},
  \bibnamefont{and} \bibinfo{author}{\bibfnamefont{A.}~\bibnamefont{Shnirman}},
  \bibinfo{journal}{unpublished}  (\bibinfo{year}{2023}).

\bibitem[{\citenamefont{Vaks et~al.}(1962)\citenamefont{Vaks, Galitskii, and
  Larkin}}]{vaks_collective_1962}
\bibinfo{author}{\bibfnamefont{V.~G.} \bibnamefont{Vaks}},
  \bibinfo{author}{\bibfnamefont{V.~M.} \bibnamefont{Galitskii}},
  \bibnamefont{and} \bibinfo{author}{\bibfnamefont{A.~I.}
  \bibnamefont{Larkin}}, \bibinfo{journal}{Soviet Physics JETP}
  \textbf{\bibinfo{volume}{14}} (\bibinfo{year}{1962}).

\bibitem[{\citenamefont{Kleinert}(1978)}]{kleinert_collective_1978}
\bibinfo{author}{\bibfnamefont{H.}~\bibnamefont{Kleinert}},
  \bibinfo{journal}{Fortschritte der Physik} \textbf{\bibinfo{volume}{26}},
  \bibinfo{pages}{565} (\bibinfo{year}{1978}),
  \urlprefix\url{https://doi.org/10.1002/prop.19780261102}.

\bibitem[{\citenamefont{Varma}(2002)}]{varma_higgs_2002}
\bibinfo{author}{\bibfnamefont{C.~M.} \bibnamefont{Varma}},
  \bibinfo{journal}{Journal of Low Temperature Physics}
  \textbf{\bibinfo{volume}{126}}, \bibinfo{pages}{901} (\bibinfo{year}{2002}),
  \urlprefix\url{https://link.springer.com/article/10.1023/A:1013890507658}.

\bibitem[{\citenamefont{Zhang et~al.}(2008)\citenamefont{Zhang, Liu, Qi, Dai,
  Fang, and Zhang}}]{zhang_topological_2008}
\bibinfo{author}{\bibfnamefont{H.}~\bibnamefont{Zhang}},
  \bibinfo{author}{\bibfnamefont{C.-X.} \bibnamefont{Liu}},
  \bibinfo{author}{\bibfnamefont{X.-L.} \bibnamefont{Qi}},
  \bibinfo{author}{\bibfnamefont{X.}~\bibnamefont{Dai}},
  \bibinfo{author}{\bibfnamefont{Z.}~\bibnamefont{Fang}}, \bibnamefont{and}
  \bibinfo{author}{\bibfnamefont{S.-C.} \bibnamefont{Zhang}},
  \emph{\bibinfo{title}{Topological {Insulators} at {Room} {Temperature}}}
  (\bibinfo{year}{2008}), \bibinfo{note}{arXiv:0812.1622 [cond-mat]},
  \urlprefix\url{http://arxiv.org/abs/0812.1622}.

\bibitem[{\citenamefont{Xia et~al.}(2008)\citenamefont{Xia, Wray, Qian, Hsieh,
  Pal, Lin, Bansil, Grauer, Hor, Cava et~al.}}]{xia_electrons_2008}
\bibinfo{author}{\bibfnamefont{Y.}~\bibnamefont{Xia}},
  \bibinfo{author}{\bibfnamefont{L.}~\bibnamefont{Wray}},
  \bibinfo{author}{\bibfnamefont{D.}~\bibnamefont{Qian}},
  \bibinfo{author}{\bibfnamefont{D.}~\bibnamefont{Hsieh}},
  \bibinfo{author}{\bibfnamefont{A.}~\bibnamefont{Pal}},
  \bibinfo{author}{\bibfnamefont{H.}~\bibnamefont{Lin}},
  \bibinfo{author}{\bibfnamefont{A.}~\bibnamefont{Bansil}},
  \bibinfo{author}{\bibfnamefont{D.}~\bibnamefont{Grauer}},
  \bibinfo{author}{\bibfnamefont{Y.~S.} \bibnamefont{Hor}},
  \bibinfo{author}{\bibfnamefont{R.~J.} \bibnamefont{Cava}},
  \bibnamefont{et~al.}, \emph{\bibinfo{title}{Electrons on the surface of
  {Bi2Se3} form a topologically-ordered two dimensional gas with a non-trivial
  {Berry}'s phase}} (\bibinfo{year}{2008}), \bibinfo{note}{arXiv:0812.2078
  [cond-mat]}, \urlprefix\url{http://arxiv.org/abs/0812.2078}.

\bibitem[{\citenamefont{Jackiw and Rebbi}(1976)}]{jackiw_solitons_1976}
\bibinfo{author}{\bibfnamefont{R.}~\bibnamefont{Jackiw}} \bibnamefont{and}
  \bibinfo{author}{\bibfnamefont{C.}~\bibnamefont{Rebbi}},
  \bibinfo{journal}{Physical Review D} \textbf{\bibinfo{volume}{13}},
  \bibinfo{pages}{3398} (\bibinfo{year}{1976}), ISSN \bibinfo{issn}{0556-2821},
  \urlprefix\url{https://link.aps.org/doi/10.1103/PhysRevD.13.3398}.

\bibitem[{\citenamefont{Grosfeld and Stern}(2011)}]{grosfeld_observing_2011}
\bibinfo{author}{\bibfnamefont{E.}~\bibnamefont{Grosfeld}} \bibnamefont{and}
  \bibinfo{author}{\bibfnamefont{A.}~\bibnamefont{Stern}},
  \bibinfo{journal}{Proceedings of the National Academy of Sciences}
  \textbf{\bibinfo{volume}{108}}, \bibinfo{pages}{11810}
  (\bibinfo{year}{2011}), ISSN \bibinfo{issn}{0027-8424, 1091-6490},
  \urlprefix\url{https://www.pnas.org/doi/10.1073/pnas.1101469108}.

\bibitem[{\citenamefont{Zakharov and
  Mikhailov}(1980)}]{zakharov_integrability_1980}
\bibinfo{author}{\bibfnamefont{V.~E.} \bibnamefont{Zakharov}} \bibnamefont{and}
  \bibinfo{author}{\bibfnamefont{A.~V.} \bibnamefont{Mikhailov}},
  \bibinfo{journal}{Communications in Mathematical Physics}
  \textbf{\bibinfo{volume}{74}}, \bibinfo{pages}{21} (\bibinfo{year}{1980}),
  ISSN \bibinfo{issn}{0010-3616, 1432-0916},
  \urlprefix\url{http://link.springer.com/10.1007/BF01197576}.

\bibitem[{\citenamefont{Brazovskiĭ}(1980)}]{brazovskii_self-localized_1980}
\bibinfo{author}{\bibfnamefont{S.~A.} \bibnamefont{Brazovskiĭ}},
  \bibinfo{journal}{Soviet Physics JETP} \textbf{\bibinfo{volume}{51}},
  \bibinfo{pages}{342} (\bibinfo{year}{1980}),
  \urlprefix\url{http://www.worldscientific.com/doi/abs/10.1142/9789814317344_0024}.

\bibitem[{\citenamefont{Takayama et~al.}(1980)\citenamefont{Takayama, Lin-Liu,
  and Maki}}]{takayama_continuum_1980}
\bibinfo{author}{\bibfnamefont{H.}~\bibnamefont{Takayama}},
  \bibinfo{author}{\bibfnamefont{Y.~R.} \bibnamefont{Lin-Liu}},
  \bibnamefont{and} \bibinfo{author}{\bibfnamefont{K.}~\bibnamefont{Maki}},
  \bibinfo{journal}{Physical Review B} \textbf{\bibinfo{volume}{21}},
  \bibinfo{pages}{2388} (\bibinfo{year}{1980}), ISSN \bibinfo{issn}{0163-1829},
  \urlprefix\url{https://link.aps.org/doi/10.1103/PhysRevB.21.2388}.

\bibitem[{\citenamefont{Takahashi and
  Nitta}(2013)}]{takahashi_self-consistent_2013}
\bibinfo{author}{\bibfnamefont{D.~A.} \bibnamefont{Takahashi}}
  \bibnamefont{and} \bibinfo{author}{\bibfnamefont{M.}~\bibnamefont{Nitta}},
  \bibinfo{journal}{Physical Review Letters} \textbf{\bibinfo{volume}{110}},
  \bibinfo{pages}{131601} (\bibinfo{year}{2013}), ISSN
  \bibinfo{issn}{0031-9007, 1079-7114},
  \urlprefix\url{https://journals.aps.org/prl/abstract/10.1103/PhysRevLett.110.131601}.

\bibitem[{\citenamefont{Takahashi}(2016)}]{takahashi_exhaustive_2016}
\bibinfo{author}{\bibfnamefont{D.~A.} \bibnamefont{Takahashi}},
  \bibinfo{journal}{Progress of Theoretical and Experimental Physics}
  \textbf{\bibinfo{volume}{2016}}, \bibinfo{pages}{043I01}
  (\bibinfo{year}{2016}), ISSN \bibinfo{issn}{2050-3911},
  \urlprefix\url{https://academic.oup.com/ptep/article/2016/4/043I01/2461117}.

\end{thebibliography}

\onecolumngrid
\appendix

\section{Deriving the low-energy BdG Hamiltonian for the SMS junction \label{app_deriving-effham}}
Starting from the Hamiltonian given in (\ref{eq_hSMS}), by adopting a semi-classical description of the magnetization and labeling the fermionic states by their momentum $-i\partial_x \rightarrow q$, which is valid if $\vec{m}$ and $\varphi$ only vary slowly with $x$, a low-energy effective Hamiltonian for the fermionic degrees of freedom in the considered SMS junction can be derived analogously to what was done in Ref.\ \onlinecite{fu_superconducting_2008}. To this end, we consider the limit where the width of the junction is much smaller than the superconducting coherence length $W\ll v_{\rm F}/\Delta_0$,  such that there exist only two branches of in-gap bound states (related by particle-hole symmetry). Splitting (\ref{eq_hSMS}) into two parts, $h=h^{(0)}+h^{(1)}$, with $h^{(0)}=h|_{q=m_x=m_y=0,\varphi=\pi}$ and treating $h^{(1)}$ as a small perturbation, one then finds zero-energy solutions $h^{(0)}\zeta_{a=1,2}=0$, onto which $h$ can be projected to define the effective BdG Hamiltonian. It is convenient to choose them such that they obey $\mathcal{C}\zeta_a=\zeta_a$ with the charge conjugation operator $\mathcal{C}=\tau_y\sigma_y\mathcal{K}$, where $\mathcal{K}$ denotes complex conjugation. They can be written as
\begin{equation}
    \begin{split}
        \zeta_\pm(y)  \equiv \zeta_1 \pm i\zeta_2(y) \propto
&\begin{bmatrix}
-\sqrt{\mu+M}
\begin{pmatrix}
-\cos\frac{\mu\chi(y)}{v_{\rm F}}\\
\sin\frac{\mu\chi(y)}{v_{\rm F}}\\
\sin\frac{\mu\chi(y)}{v_{\rm F}}\\
\cos\frac{\mu\chi(y)}{v_{\rm F}}
\end{pmatrix}
\pm i\sqrt{\mu-M}
\begin{pmatrix}
\sin\frac{\mu\chi(y)}{v_{\rm F}}\\
\cos\frac{\mu\chi(y)}{v_{\rm F}}\\
\cos\frac{\mu\chi(y)}{v_{\rm F}}\\
-\sin\frac{\mu\chi(y)}{v_{\rm F}}
\end{pmatrix} 
\end{bmatrix}\times\\
&\hspace{0.5cm}\times\exp\left[\pm i\sqrt{\mu^2-M^2}(y-\chi(y))/v_{\rm F}-\int_0^{|y|}\text{d}\tilde{y}\,\Delta_0(\tilde{y})/v_{\rm F}\right].
    \end{split}
\end{equation}
The normalization is chosen such that $\braket{\zeta_a|\zeta_a}=\int_{-\infty}^\infty\text{d}y\,|\zeta_a(y)|^2=1/L$ holds, from which $|C_{1/2}|^2\simeq\frac{\Delta_0}{2v_{\rm F}L}\frac{1}{\mu\pm MK}$ follows. Additionally we defined $\chi(y)=\left(y-\text{sgn}(y)\frac{W}{2}\right)\Theta\left(|y|-\frac{W}{2}\right)$. Small values of $q$, $m_x$, $m_y$ and $\pi-\varphi$ are now included by calculating the matrix elements
\begin{equation}
\begin{split}
\braket{\zeta_a|\Delta_0(\pi-\varphi)\Theta(y-W/2)\tau_y|\zeta_b} &= \sqrt{\frac{\mu^2-M^2}{\mu^2-M^2K^2}}\frac{\Delta_0(\pi-\varphi)}{2L}\tilde{\tau}_y^{ab},\\
\braket{\zeta_a|v_\text{F}q\tau_z\sigma_x|\zeta_b} &= \frac{vq}{L}\tilde{\tau}_x^{ab},\\
\braket{\zeta_a|M\Theta(W/2-|y|)m_x\sigma_x|\zeta_b} &= \braket{\zeta_a|M\Theta(W/2-|y|)\frac{m_x^2+m_y^2}{2}\sigma_z|\zeta_b} = 0, \\
\braket{\zeta_a|M\Theta(W/2-|y|)m_y\sigma_y|\zeta_b} &= \sqrt{\frac{\mu^2-M^2}{\mu^2-M^2K^2}}\frac{\Delta_0 MW}{v_\text{F}L}m_y\tilde{\tau}_y^{ab},
\end{split}
\end{equation}
where the Pauli matrices $\tilde{\tau}_{x,y,z}$ act on $(\zeta_1,\zeta_2)$, the effective velocity reads $v=\sqrt{\frac{\mu^2-M^2}{\mu^2-M^2K^2}}\frac{\Delta_0^2}{\Delta_0^2+\mu^2}Kv_{\rm F}$ and the dimensionless constant $K= \cos\left(\sqrt{\mu^2-M^2}W/v_{\rm F}\right) + \frac{\Delta_0}{\sqrt{\mu^2-M^2}}\sin\left(\sqrt{\mu^2-M^2}W/v_{\rm F}\right)$ has been defined. With that, the low-energy effective BdG Hamiltonian
\begin{equation}
    h_{\rm eff}^{ab} = \braket{\zeta_a|h|\zeta_b} = -iv\partial_x\tilde{\tau}_x^{ab}/L+\sqrt{\frac{\mu^2-M^2}{\mu^2-M^2K^2}}\Delta_0\left(\frac{\pi-\varphi}{2}+\frac{MW}{v_{\rm F}}m_y\right)\tilde{\tau}_y^{ab}/L \label{eq_heff}
\end{equation}
is obtained. We find the component of the magnetization perpendicular to the junction $m_y$ to play a similar role as the deviation of the phase difference from $\pi$. In contrast to $m_y$, the $x$-component of the magnetization $m_x$ does not directly couple to the fermionic degrees of freedom. 

Finally, by fixing $\varphi=\pi$ and introducing the right- and left-moving Majorana fields $\gamma_{\rm R/L}(x)=\frac{1}{\sqrt{L}}\sum_q \gamma_q^{\rm R/L}e^{iqx}=\gamma_{\rm R/L}^\dagger(x)$ with $\gamma^{\rm R/L}_q = \int\text{d}\bm{r}\,(\xi^{\rm R/L}_q(\bm{r}))^\dagger\Psi(\bm{r}) = (\gamma^{\rm R/L}_{-q})^\dagger$, where $\xi^{\rm R/L}_q(\bm{r})=\frac{1}{\sqrt{2}}(\zeta_1(y)\pm\zeta_2(y))e^{iqx}$, we obtain
\begin{equation}
    H_{\rm eff} = \int{\rm{d}}x\,\Big[-\frac{iv}{2}(\gamma_{\rm R}\partial_x\gamma_{\rm R}-\gamma_{\rm L}\partial_x\gamma_{\rm L})-i\lambda m_y\gamma_{\rm L}\gamma_{\rm R}\Big]
\end{equation}
with the coupling constant $\lambda = M\sqrt{\frac{\mu^2-M^2}{\mu^2-M^2K^2}}\frac{\Delta_0 W}{v_{\rm F}}$.


\end{document}